\documentstyle[12pt,aasms]{article}

\newcommand{\mincir}{\raise -2.truept\hbox{\rlap{\hbox{$\sim$}}\raise5.truept
\hbox{$<$}\ }}
\newcommand{\magcir}{\raise -2.truept\hbox{\rlap{\hbox{$\sim$}}\raise5.truept
\hbox{$>$}\ }}
\newcommand{\minmag}{\raise-2.truept\hbox{\rlap{\hbox{$<$}}\raise 6.truept\hbox
{$>$}\ }}
\newcommand{\be}{\begin{equation}}
\newcommand{\ee}{\end{equation}}
\newcommand{\ba}{\begin{eqnarray}}
\newcommand{\ea}{\end{eqnarray}}
\newcommand{\brr}{\begin{array}}
\newcommand{\err}{\end{array}}
\newcommand{\bc}{\begin{center}}
\newcommand{\ec}{\end{center}}

\newcommand{\hm}{\,{h^{-1}{\rm Mpc}}}
\newcommand{\vel}{\,{\rm km\,s^{-1}}}

\begin{document}
\baselineskip 0.7cm

\begin{titlepage}

%  \begin{flushright}
%    TUM-HEP-248/96
%    \\
%    SFB-375/95
%    \\
%   April 1996
%  \end{flushright}

\begin{center}
  {\Large \bf Peculiar Velocities of Clusters in CDM Models} \\
{\large Stefano Borgani$^{1)}$, Luiz N. da Costa$^{2,3)}$, \\
Wolfram Freudling$^{4)}$ , Riccardo Giovanelli$^{5)}$, 
Martha P. Haynes$^{5)}$,\\
John Salzer$^{6)}$ and Gary Wegner$^{7)}$
}
\vskip 0.4cm 
{\it 1) INFN, Sezione di Perugia,
    c/o Dipartimento di Fisica dell' Universit\`{a} \\
    via A. Pascoli, I-06100 Perugia, Italy} \\

{\it 2) European Southern Observatory, Karl--Schwarzschild--Str. 2, \\
    D-85748 Garching b. M\"unchen, Germany} \\

{\it 3) Observat\'orio Nacional, Rua Gen. Jos\'e Cristino 77, \\
     Rio de Janeiro, RJ, Brazil} \\

{\it 4) Space-Telescope European Coordinating Facility and European Southern Observatory, \\
    Karl--Schwarzschild--Str. 2, D-85748 Garching b. M\"unchen, Germany} \\

{\it 5) Center for Radiophysics and Space Research, and National Astronomy
     and Ionosphere Center\footnote{The National Astronomy and
     Ionosphere Center is operated by Cornell University under a
     cooperative agreement with the National Science Foundation.}, \\
     Cornell University, Ithaca, NY 14853, U.S.A.} \\

{\it 6) Astronomy Dept., Wesleyan University, Middletown, CT 06459, U.S.A.} \\

{\it 7) Department of Physics and Astronomy, Dartmouth College, Hanover,
     NH 03755, U.S.A. }\\

\vskip 0.5in

\abstract Recently, peculiar velocity measurements became available
for a new sample of galaxy clusters, hereafter the SCI sample. From an
accurately calibrated Tully--Fisher relation for spiral galaxies, we
compute the rms peculiar velocity, $V_{rms}$, and compare it to the
linear theory predictions of COBE--normalized low--density and open
CDM models ($\Lambda$CDM and OCDM, respectively). Confidence levels
for model rejection are estimated using a Monte Carlo procedure to
generate for each model a large ensemble of artificial data sets.
Following Zaroubi et al. (1997), we express our results in terms of
constraints on the $(\Omega_0,n_{pr},h)$ parameter space.  Such
constraints turn into $\sigma_8 \Omega_0^{0.6}=0.50^{+0.25}_ {-0.14}$
at the 90$\%$ c.l., thus in agreement with results from cluster
abundance. We show that our constraints are also consistent with those
implied by the shape of the galaxy power spectrum within a rather wide
range for the values of the model parameters. Finally, we point out
that our findings disagree at about the $3\sigma$ level with respect
to those by Zaroubi et al. (1997), based on the Mark III catalogue,
which tend to prefer larger $\Omega_0$ values within the CDM class of
models.
\end{center}

\end{titlepage}

\section{Introduction}

Peculiar velocities of clusters have recently been used by several 
authors to set stringent constraints on cosmological models (e.g., 
Croft \& Efstathiou 1994; Cen, Bahcall \& Gramann  1994; Bahcall \& Oh
1996; Moscardini et al. 1996). Although clusters sample the
large--scale flows much more sparsely than galaxies, their peculiar
velocity can be measured more accurately if distances are available for
several cluster galaxies. Several of such previous analyses were,
however, based on non--homogeneous compilations, with cluster
velocities taken from different parent samples.

In this {\em Letter} we analyze the new sample of cluster peculiar
velocities described by Giovanelli et al. (1997a; SCI hereafter),
consisting of accurate and uniform I-band photometry and velocity width
measurements for about 800 spiral galaxies in the fields of 24
clusters. Of the 24 clusters, we consider 18,
pruning the six paired clusters
(A2197/A2199, S805=Pavo II/Pavo, and A2634/A2666)
in order to avoid possible ambiguities in membership assignment
(see Giovanelli et al. 1997a).  

Peculiar velocities for these clusters were
determined from a a well-calibrated I-band Tully--Fisher relation
constructed from all 24 clusters as described in Giovanelli et al.
(1996b).  An earlier version of this sample has been already analyzed by
 Bahcall \& Oh (1996) and Moscardini et al. (1996), who compared the
results  with numerical simulations of several CDM--like models. Within
this class  of models, both analyses consistently showed that this data
set favors a low--density Universe, with $0.2\mincir
\Omega_0\mincir 0.4$. Furthermore, Moscardini et al. (1996)  also
compared the SCI sample with Hudson's (1994) compilation of 
cluster velocities.
They found that the SCI sample provides systematically smaller 
velocities than those of Hudson, 
again suggesting a low value of $\Omega_0$. 

The analysis that we present in this {\em Letter} is entirely based on linear
theory, the reliability of which 
to describe cluster motions is briefly discussed.
Model predictions are worked out for purely CDM models with $\Omega_0\le 1$
and both flat and open geometry. Avoiding the need
to resort to numerical simulations
allows us to probe the model parameter space in a much more accurate way.
The resulting constraints on the CDM models are also compared with those 
derived from the cluster abundances, the galaxy power--spectrum shape and the
Mark III data as analyzed by Zaroubi et al. (1997)

\section{The analysis}
Our analysis is based on comparing the rms cluster velocity, $V_{rms}$ from
the SCI sample and for models. Its observational
estimate for the SCI clusters gives a one dimensional
$V^{obs}_{rms}=266\pm 30\vel$, where the uncertainty represents the $1\sigma$ 
scatter over $10^5$ Montecarlo realizations of the real sample, each one 
generated from an {\em a priori} Gaussian distribution 
having the same $V_{rms}$
as the SCI data set and velocities convolved with the observational errors.

Linear gravitational instability predicts the
one--dimensional rms velocity to be
\be 
V_{rms}\,=\,{H_0f(\Omega_0)\over
\sqrt{3}}\,\left[{1\over 2\pi^2}\int_0^\infty
dk\,P(k)\,W^2(kR)\right]^{1/2}\,, 
\label{eq:lt} 
\ee
where $f(\Omega_0)\simeq \Omega_0^{0.6}$, $P(k)$ is the model power spectrum 
and $W(kR)$ is the window function that specifies the ``shape'' and the 
size $R$ of the linear density fluctuations, which generate clusters. 
Bahcall, Gramann \& Cen (1994) and Croft \& Efstathiou (1995) verified
that eq.(\ref{eq:lt}) provides a rather good fit to the cluster rms
velocity  generated by N--body simulations. By using the Gaussian window
$W(kR)=\exp(-k^2R^2/2)$, we found that eq.(\ref{eq:lt}) provides the best 
fit to the Borgani et al. (1997) N--body outputs for a variety of
models by taking $R=3.9\hm$ ($H_0=100h \vel$ Mpc$^{-1}$), 
corresponding to a typical cluster mass 
$M_{cl}\simeq 2.6\times 10^{14}\Omega_0 
h^{-1}M_\odot$. We adopt this value
in the following analysis. 

We take the power spectrum to be
$P(k)=Ak^{n_{pr}}T^2(k)$,  where \be T(q)\,=\,{\ln(1+2.34q)\over
2.34q}\,\left[1+3.89q+(16.1q)^2+(5.46q)^3+ (6.71q)^4\right]^{-1/4}
\label{eq:tk} \ee is the CDM transfer function provided by Bardeen et
al. (1986). Here $q=k/\Gamma h$ where $\Gamma=\Omega_0
h\exp(-\Omega_b- (2h)^{1/2}\Omega_b/\Omega_0)$ is the ``shape''
parameter, which takes into account the presence of a non--negligible
baryon fraction, $\Omega_b$ (e.g., Sugiyama 1995). We take
$\Omega_bh^2= 0.024$ (e.g., Tytler et al. 1995). The constant $A$ is
fixed by the 4--year {\sl COBE} normalization recipes of Bunn \& White
(1996; see their eqs. 29 and 31)
and Hu \& White (1997; see their eq. 6)
for flat low--density ($\Lambda$CDM) and open (OCDM) models
for both vanishing and non--vanishing tensor mode contributions to CMB
anisotropies.  In the case of $\Lambda$CDM, Bunn \& White (1996)
consider the case $T/S=7(1-n_{pr})$ for the ratio betweem the
quadrupole moments of the tensor and scalar modes in the expansion of
the angular temperature fluctuations as generated by power--law
inflation (e.g., Crittenden et al. 1993, and references therein). As
for OCDM, the normalization for the $T/S\ne 0$ case is provided by Hu
\& White (1997) in the case of ``minimal'' tensor anisotropies.
The density parameter $\Omega_0$ and the primordial
spectral index $n_{pr}$ are varied within the ranges where the
normalization fits
are reliable, namely $0.2\le  \Omega_0 \le 1$ and $0.7 \le
n_{pr} \le 1.2$ 
($0.7 \le n_{pr} \le 1$) for the cases without (with) a tensor mode
contribution.

The family of models we consider is specified by the three 
parameters $(\Omega_0,n_{pr},h)$. Results will be presented by 
keeping one of them fixed, in the form of slices of the
three--dimensional parameter space.
Colberg et al. (1997) have recently pointed out that cluster peculiar
velocities are almost independent of the density parameter, once
cosmological models are normalized to reproduce the cluster abundance,
according to the recipe by Eke et al. (1996). Here we follow the
different approach of imposing the COBE normalization, since we regard
CMB temperature anisotropies as a more stable and robust constraint
than cluster abundance for a fixed choice of model parameters.

In order to establish the confidence level for the validity of a given 
model, we adopt the following procedure. Let $v_i$ and $\sigma_i$ 
be the velocity and its error for the $i$-th real cluster ($i=1,\dots,18$).
For a given model, we generate Montecarlo samples, each containing 18 
velocities, $V_i$, drawn from a Gaussian
distribution, having dispersion given by eq.(\ref{eq:lt}).
For each sample we convolve the $i$--th model velocity with the 
observational error $\sigma_i$ and estimate 
the resulting rms velocity. 
For each sample, every cluster's velocity is newly estimated as a
Gaussian deviate of the mean $V_{i}$ and dispersion $\sigma_{i}$, and
the rms velocity of the sample then computed.

For each model we generate $N=10^4$ samples and then compute the
fraction ${\cal F}$ of them with $V^j_{rms}$ ($j=1,\dots,N$) 
at least as 
discrepant as $V^{obs}_{rms}$ with respect to their average value, 
$N^{-1}\sum_j V^j_{rms}$:
the smaller the value of ${\cal F}$, the smaller the probability that
$V^{obs}_{rms}$ is generated by chance by that model, 
the larger the probability ${\cal P}=1-{\cal F}$ 
that the model itself is rejected.

Figure 1 shows the results of our analysis for scale--free (i.e.,
$n_{pr}=1$) $\Lambda$CDM and OCDM models, also comparing them to other
observational constraints. The contours indicate the iso--probability
levels for the model exclusion. The outermost contour is for ${\cal
  P}=90\%$ confidence level, while different levels are equi--spaced
in logarithmic units by $\Delta(\log{\cal P}) =0.1$. The heavily
shaded area indicates the $1\sigma$ confidence level from the cluster
abundance by Eke et al. (1996). By fitting the X--ray cluster
temperature function with CDM model predictions, they found
$\sigma_8\Omega_0^\alpha =0.52\pm 0.04$, with
$\alpha=0.52-0.13\Omega_0$ for $\Lambda$CDM and
$\alpha=0.46-0.10\Omega_0$ for OCDM ($\sigma_8$ is the linear rms
density fluctuation within a top--hat sphere of $8\hm$). \footnote{Pen
  (1996) recently pointed out that the scaling by Eke et al. (1996)
  somewhat underestimates the value of $\sigma_8$ at small $\Omega_0$
  values. In particular, he claimed that for a $\Lambda$CDM model with
  $\Omega_0\simeq 0.35$ $\sigma_8$ should be $\sim 17\%$ larger. We
  checked that the central value for the $h$ interval corresponding to
  such an $\Omega_0$ increase from $h=0.66$ to $h=0.74$.} The
medium--weight
shaded area is the 95$\%$ confidence level 
from the fitting by Liddle et al. (1996)
to the shape of the APM galaxy power spectrum by Peacock \& Dodds 
(1994): $\Gamma=0.23-0.28(1-1/n_{pr})$ with errors of about $16\%$. A 
consistent result has also been found by Borgani et al. (1997) from the 
analysis of the cluster distribution. The lightly
shaded area shows the 90$\%$ confidence level by Zaroubi et al. (1997;
Z97 hereafter) from the likelihood analysis of the Mark III galaxy
peculiar velocities (Willick et al. 1996), whose results are reported
in Table 1. The dashed curves are for different values for the age of
the Universe: $t_0=9,11,13,15,17$ Gyrs from upper to lower curves.

Our results differ with respect to those by Z97.
The difference is larger for $\Lambda$CDM models, for which the
discrepancy is at $\sim 3\sigma$ level
(note that the corresponding 90$\%$ confidence level are at most
marginally overlapping), the latter favoring larger $\Omega_0$ values
(at a fixed $h$). This result points in the same direction as that
found by Moscardini et al. (1996). On the other hand, the constraints
we set on CDM models are quite consistent with those coming from the
$P(k)$ shape and the cluster abundance, in a rather broad range of
$\Omega_0$ and $h$ values. For instance, if we impose ages in the
range $13 \mincir t_0\mincir 15$ Gyrs, $\Lambda$CDM models require
$0.35\mincir \Omega_0\mincir 0.50$ with $0.50\mincir h\mincir 0.65$,
while OCDM models require $0.50\mincir \Omega_0\mincir 0.70$ with
$0.45\mincir h\mincir 0.60$. On the contrary, the results by Z97 are
rather discrepant with both such constraints, especially with the
cluster abundance.

A similar picture also emerges as tilted (i.e., $n_{pr}\neq 1$) models are
considered. 
Figure 2 is analogous to Fig.1 , but with results plotted in the
$n_{pr}$--$\Omega_0$ plane, taking $h=0.65$ and 0.55 for $\Lambda$CDM
and OCDM models, respectively.  For both classes of models, taking
$T/S\ne 0$ has the effect of
{\em (a)} narrowing the permitted 
region in the parameter space and {\em (b)} decreasing the need for a tilt
(cf. also Z97). Tilting $P(k)$ breaks
the degeneracy of the $P(k)$ shape with the other constraints.
Fixing $h=0.65$ (e.g. Giovanelli et al. 1997c) and $t_0\simeq 13$ Gyrs 
for $\Lambda$CDM would require $\Omega_0= 0.43$; this turns into
$0.85\mincir n_{pr}\mincir 0.95$ and $0.90\mincir n_{pr}\mincir 0.96$ for
$T/S=0$ and $T/S=7(1-n_{pr})$, respectively.
Consistency between the $P(k)$ shape and Z97 are attained for
$n_{pr}\magcir 1$ and $\Omega_0\mincir 0.5$, while the cluster abundance is
still largely missed. As for OCDM models,
taking $h=0.55$ and $t_0\simeq 13$ Gyrs implies $\Omega_0\simeq 0.65$
and $0.84\mincir n_{pr}\mincir 0.94$ ($0.84\mincir n_{pr}\mincir
0.94$) for $T/S=0$ ($\ne 0$).
A substantially larger $h$ value would turn into too small $\Omega_0$
values, unless $t_0< 13$ Gyrs. Again, the SCI cluster velocities are
consistent in all the cases with the other two constraints for
reasonable values of the model parameters.

In order to better quantify the difference with respect to the
constraints provided by the Z97 analysis, we fit the same combination
of parameters, $\Omega_0 h_{50}^\mu n_{pr}^\nu=C$ ($h_{50}=2\,h$:
Hubble constant in units of 50$\vel$Mpc$^{-1}$), considered in that paper and
the results are in Table 1.  Although the shape of the relation (i.e.,
the values of $\mu$ and $\nu$) is quite similar, its amplitude $C$ is
significantly different. This confirms that, for fixed $h$ and
$n_{pr}$ values, our results favor a lower density parameter.  We have
also computed the best fit to the quantity $\sigma_8 \Omega_0^{0.6}$,
which fixes the amplitude of the velocity field in linear theory. We
find $\sigma_8 \Omega_0^{0.6} =0.50^{+0.25}_{-0.17}$ (errors
correspond to 90$\%$ confidence level), which again agrees with the
constraints from the cluster abundance, and is significantly smaller
than $\sigma_8 \Omega_0^{0.6}=0.88\pm 0.15$ that derived by Z97.
Willick et al. (1996) recently compared the Mark III data with
velocity and density fields reconstructed from the 1.2 Jy {\sl IRAS}
survey. Quite interesetingly, they obtained $\sigma_8\Omega_0^{0.6} =
0.34 \pm 0.05$ (cf. their Fig. 20), thus at variance with respect to
the results by Z96, also based on the Mark III sample, and rather
consistent with our results.

\section{Conclusions}

We have performed a detailed comparison of the cluster peculiar
velocities in the SCI catalog with those
predicted by COBE--normalized CDM models, using linear theory. This
comparison has been made by computing the rms cluster velocity,
$V_{rms}$, for data and models, and estimating the likelihood that the
observed value $V_{rms}=266\pm 30 \vel$ is consistent with a given
model.

Confidence levels for rejecting models were determined using a
Monte Carlo procedure which generates a large number ($10^4$) of mock
samples from each model. The main goal of our analysis has been to
impose constraints on the space of $(\Omega_0,n_{pr},h)$ parameters for
CDM models. We have compared our results with those of Z97, and with the
constraints that have been established from the properties of
clustering of galaxies as expressed by the shape of the power--spectrum
and recent determinations of cluster abundance.

Our results can be summarized as follows:

\begin{description}
\item[(a)] Velocities of SCI clusters point toward a low--normalization model,
characterized by $\sigma_8 \Omega_0^{0.6}=0.50^{+0.25}_{-0.17}$. This result
agrees with the independent constraint coming from the abundance 
of galaxy clusters. 

\item[(b)] Our results disagree at about the $3\sigma$ level with 
those of Z97,
based on Mark III,
the latter generally indicating higher velocities and, therefore,
favoring larger $\Omega_0$ values for fixed $h$ and $n_{pr}$
parameters (cf. Table 1). 
On the other hand, we are quite consistent with the analysis by
Willick et al. (1996), which is also based on the Mark III sample.

\item[(c)] The results agree well with 
those from the analysis of field spirals in 
the new SFI sample (da Costa et al. 1997; Freudling et al. 1997).

\end{description}

The conclusions that we draw in this {\em Letter} about the values of
the model parameters strictly hold, only for the CDM class of models.
For instance, Cold+Hot DM models, are characterized by different power
spectrum shapes and smaller COBE--normalized $\sigma_8$ values for a
fixed choice of $(\Omega_0,n_{pr},h)$, depending on the amount and the
nature of the hot component (e.g., Primack 1996, and references
therein). We postpone to a forthcoming paper the analysis of a wider
class of cosmological models, as well as the comparison with other
data sets for galaxy and cluster peculiar velocities.

\section*{Acknowledgments}
SB wishes to acknowledge ESO in Garching and SISSA in Trieste for their 
hospitality during the preparation of this work. This work was
supported by NSF grants AST94-20505 to RG, AST90-14860 and AST9023450
to MH and AST93-47714 to GW who also received partial support from 
ESO in Garching.

%\newpage

%\noindent

\newpage
\section*{Figure captions}

\noindent
{\bf Figure 1.} Constraints from the SC cluster sample on the
$(h,\Omega_0)$ plane for scale--free ($n_{pr}=1$) models. The
countours are the levels at equal probability ${\cal P}$ for model
rejection. The most external level corresponds to ${\cal P}=90\%$ and
the spacing corresponds to $\Delta(\log {\cal P})=0.2$. The heavely
shaded area is the constraint from cluster abundance (after Eke et al.
1996), the medium--weight shaded area is for the shape of the APM
galaxy power spectrum (after Viana et al. 1996) and the lightly shaded
area is from the analysis of Mark III velocities by Zaroubi et al.
(1996). Dashed curves indicate different ages for the Universe:
$t_0=9,11,13,15,17$ from upper to lower curves.

\noindent
{\bf Figure 2.} It is analogous to Fig. 1, but on the $(n_{pr},\Omega_0)$ 
plane. Different ages of the Universe are now indicated with the vertical 
dashed lines. 

\newpage
\begin{table}[tp]
\centering
\caption[]{Values of thw fitting parameters for the relation 
$\Omega_0h_{50}^\mu n_{pr}^\nu=C$ ($h_{50}=2\,h$: Hubble constant in
units of 50$\vel$Mpc$^{-1}$), from our analysis and from that by Zaroubi 
et al. (1997; Z97).}
\tabcolsep 5pt
\begin{tabular}{lcccccc} \\ \\ \hline \hline
Model & \multicolumn{3}{c}{This paper} & \multicolumn{3}{c}{Z97} \\
 & $\mu$ & $\nu$ & $C$             & $\mu$ & $\nu$ & $C$ \\ \hline
$\Lambda$CDM  $T/S=0$           & 1.30 &$1.8^{+0.2}_{-0.4}$ 
&$0.53^{+0.17}_{-0.14}$ &1.30 &2.0 &$0.83\pm 0.12$ \\
$\Lambda$CDM  $T/S=7(1-n_{pr})$ & 1.30 &$3.2^{+0.4}_{-0.7}$ 
&$0.53^{+0.17}_{-0.14}$ &0.87 &3.4 &$0.83\pm 0.12$ \\
OCDM  $T/S=0$                   & 0.87 &$1.3^{+0.1}_{-0.2}$ 
&$0.67^{+0.15}_{-0.14}$ &0.95 &1.4 &$0.88\pm 0.09$ \\ 
OCDM  $T/S\ne 0$                & 0.87 &$2.2^{+0.2}_{-0.3}$ 
&$0.67^{+0.15}_{-0.14}$ & & & \\ \hline

\end{tabular}
\label{t:fit}
\end{table}

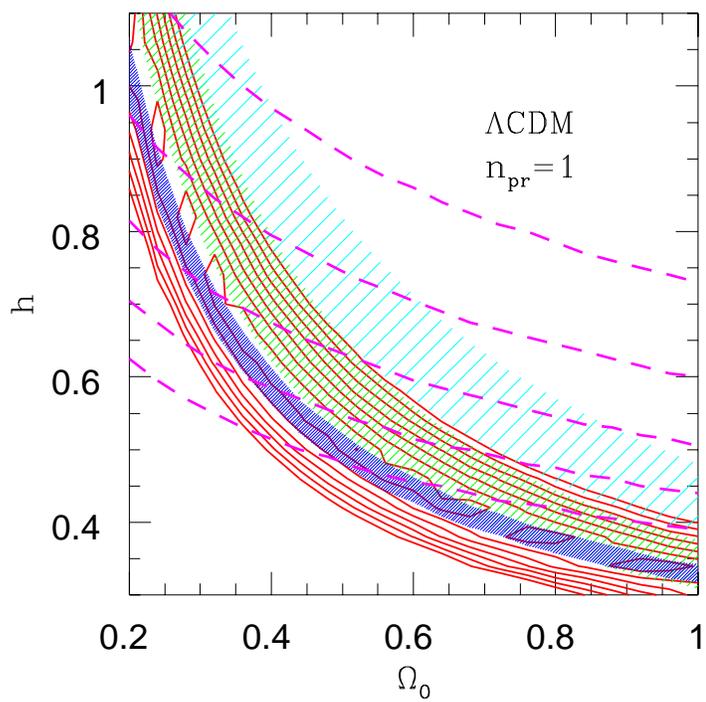
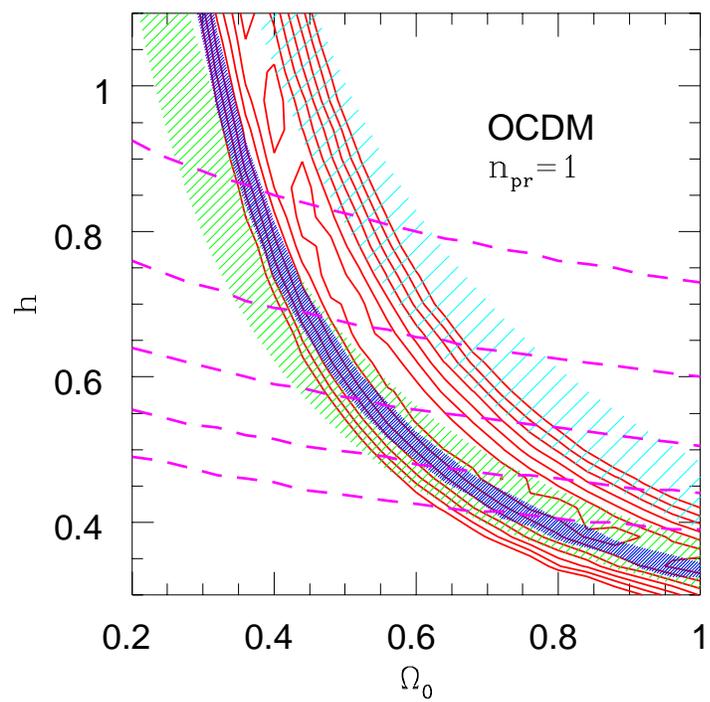
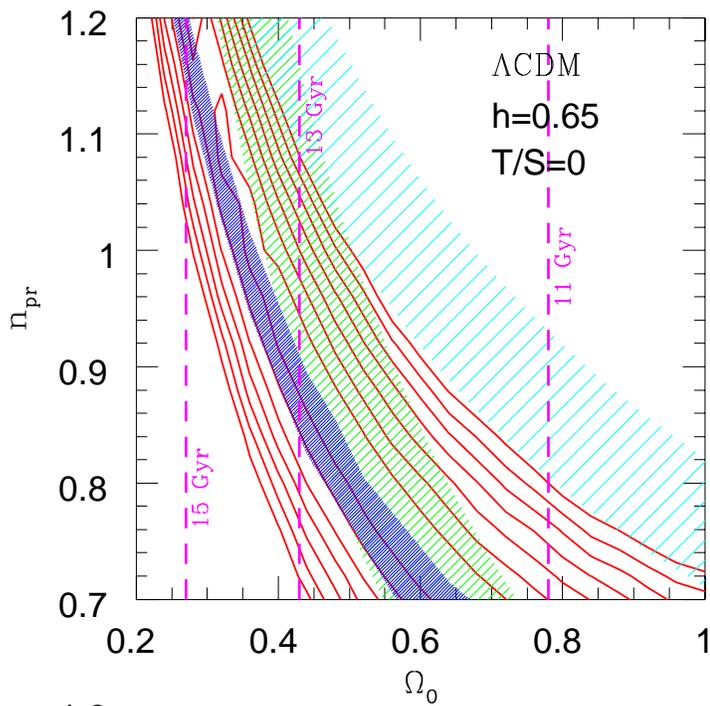
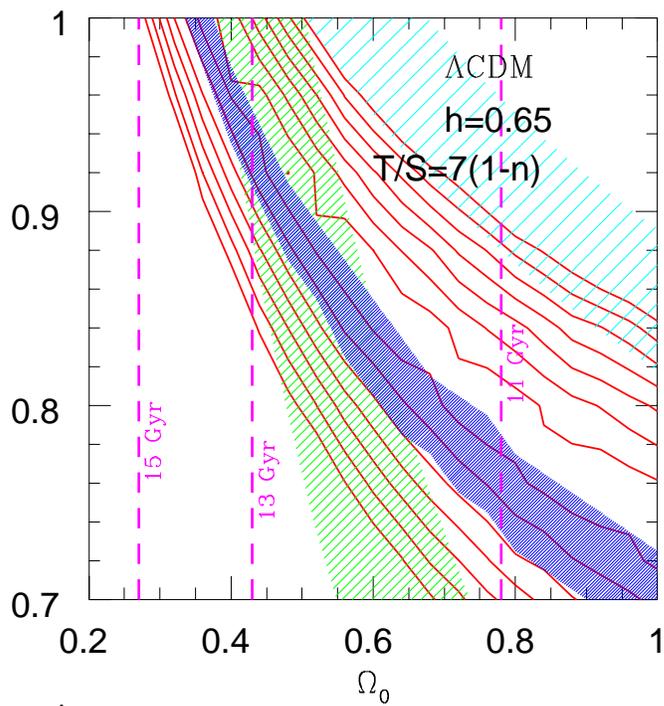
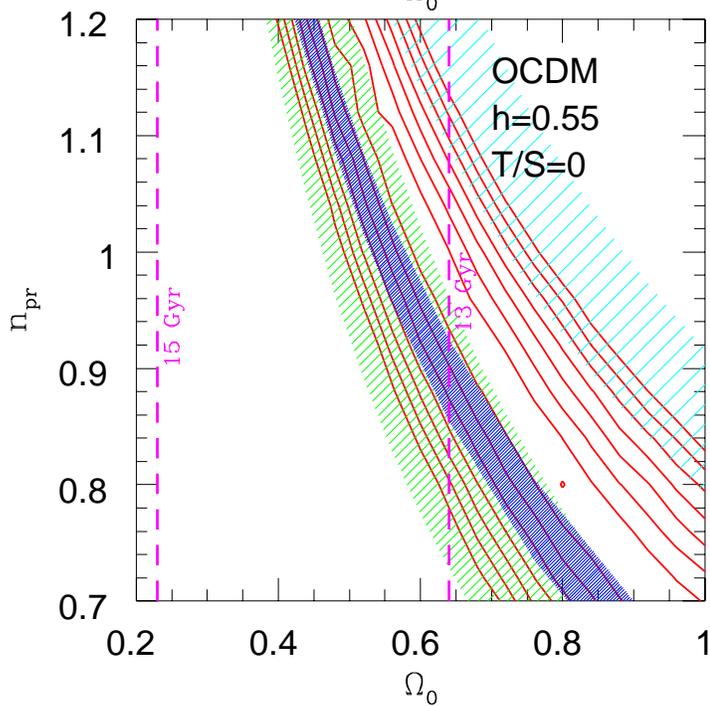
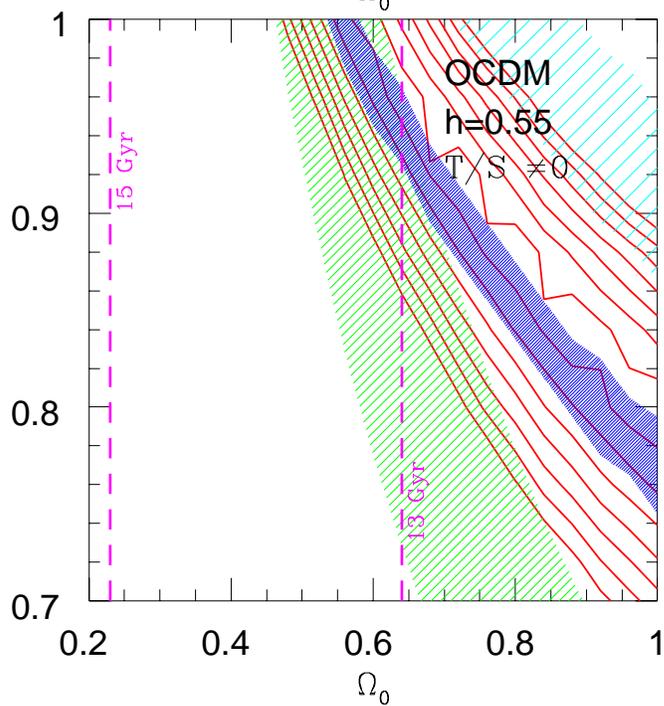
\end{document}